\documentclass{article}
\begin{document}
\begin{center}
y-DEFORMED BPS Dp- BRANES ON A SURFACE  IN A CALABI-YAU THREEFOLD\\ [.25in]
by Juan F. Ospina G.
\end{center}
\begin{center}
ABSTRACT \\ [.25in]
Using y-deformed algebraic geometric techniques the y-deformed Mukay vector of RR-charges of the y-deformed BPS Dp-branes localized on a sufarce in a Calabi-Yau threefold. The formules that are obtained here are generalizations of the formulaes of the fourth section of the preprint hep-th/0007243
\end{center}

\section{Introduction:y-deformed BPS Dp-branes on a Calabi-Yau threefold}
\setlength{\baselineskip}{20pt}

A BPS D-brane on a Calabi-Yau threefold X can be represented using a  coherent $O_{X}$-module G. The RR charge of G is given by the Mukai vector[1]:
\begin{center}
{ \mathversion{bold} $ v_{X}(G)=ch(G)\sqrt{Todd(T_{X})}{\in}H_{2*}(X;Q):={\oplus}_{i=0}^3H_{2i}(X;Q)  $ }
\end{center}

where $ ch(G)= \sum_{i=0}^{3}ch_{i}(G)$ is the Chern character with $ ch_{i}(G){\in}H_{6-2i}(X;Q)$, which can be computed by the homology-cohomology duality[1]:  always one can to have a resolution of G by locally free sheaves  $(V_{*}) $,in such way that one can to set that
$ ch(G):= \sum_{i=0}^{3}(-1)^ich(V_i)$,and these result does not depend on the choise of the resolution. Finally  $Todd(T_{X})=[X]+\frac{\rm c_{1}[X]}{\rm 2}+\frac{\rm c_{2}[X]+c_{1}[X]^2}{\rm 12}+\frac{\rm c_{2}[X]c_{1}[X]}{\rm 24} $. Now when X is a Calabi-Yau threefold one has $c_{1}[X]=0
   $ and then one obtains: $Todd(T_{X})=[X]+\frac{\rm c_{2}[X]}{\rm 12}$.
From these the effect of the square root of the Todd Class on the RR charges,is to say the geometric version of the Witten effect is given by:
\begin{center}
{ \mathversion{bold} $\sqrt{Todd(T_{X})} =[X]+\frac{\rm c_{2}[X]}{\rm 24} $ }
\end{center}

For the investigation of the topological aspects of D-branes is of the great importance to obtain several basic invariants of BPS D-Branes. One of these invariants is the RR charge of the D-brane. Other invariant is the intersection form on D-branes on X [1]. This invariant for intersections of 
two Dp-branes is obtained by multiplication of the Mukay vectors of RR charges corresponding to the intersecting Dp-branes and is given by: [1]
 
\begin{center}
\setlength{\baselineskip}{30pt}
{ \mathversion{bold} $ I_{X}(G_1,G_2)=[v_{X}(G_1)^v.v_{X}(G_2)]_X=[(ch(G_1)\sqrt{Todd(T_{X})})^v.ch(G_2)\sqrt{Todd(T_{X})}]_X=[ch(G_1)^v.ch(G_2)Todd(T_{X})]_X $ }
\end{center}
where $ [...]_X $  evaluates the degree of $ H_{0}(X;Q){\cong}Q$ component, and $v^{\vee}$ flips the sign of $H_{0}(X){\oplus}H_{4}(X)$ part of the Mukay vector v . In particular, if G itself is locally free, then $ch(G)^{\vee}=ch(G^{\vee})$, where  $ G^{\vee}=Hom_{X}(G,0_X)$ is the dual sheaf. Finally is easy to check that: $I_{X}(G_1,G_2) =-I_{X}(G_2,G_1)$.
On other hand the invariant of intersection between D-branes is an application of the Hirzebruch-Riemann-Roch and for then you can write[1]
\begin{center}

{ \mathversion{bold} $ I_{X}(G_1,G_2)=\sum_{i=0}^{3}(-1)^idimExt^i_X(G_1,G_2) $ }
\end{center}
For this reason the skew-symmetric property $I_{X}(G_1,G_2) =-I_{X}(G_2,G_1)$ of the intersection form $I_X$  for the intersection of two Dp-branes may be attributed to the Serre duality: $Ext^i_X(G_1,G_2){\cong}Ext^{3-i}_X(G_1,G_2)^{\vee}$ [1]. Another interesting comentary is that from the integrality theorems for diferential and complex manifolds the formula H.R.R. is an integer and this assures that
$I_X$ takes values in Z. [1],[2].
 
Now the result that this work presents is about the y-deformed Dp-branes on a Calabi Yau threefold. A y-deformed BPS Dp-brane on a Calabi-yau X can be represented by a y-deformed  $O_{X}-modulo G$.  The y-deformed RR charge of G is given by the y-deformed Mukai vector:
\begin{center}
{ \mathversion{bold} $ v_{X,y}(G)=ch_y(G)\sqrt{{\chi}_y(T_{X})}{\in}(H_{2*}(X;Q){\otimes}Q[y]):={\oplus}_{i=0}^3(H_{2i}(X;Q){\otimes}Q[y])  $ }
\end{center}
where ${\chi}_y$ is the y-chi-genus which is a generalization of the Todd class [2,3] and $ch_y(G)$ is the y-deformed Chern Character. the total Chern Class for $T_{X}$  has the following sumarization:
\begin{center}
{ \mathversion{bold} $ c(T_{X}) = \sum_{j=0}^{3}c_j(T_{X}) $ }
\end{center} 
also, the total Chern Class for the such bundle has the following factorization:

\begin{center}
{ \mathversion{bold} $ c(T_{X}) = \prod_{i=1}^{3}(1+x_i)$ }
\end{center}
The  CHI-y- genus for $T_{X}$ has the following formal factorisation:
\begin{center}
{ \mathversion{bold} $ {\chi}_y(T_{X}) = \prod_{i=1}^3\frac{\rm(1+yexp(-(y+1)x_i))x_i }{\rm 1-exp(-(y+1)x_i)}$ }
\end{center}

The CHI-y- genus for $T_{X}$  has the following formal sumarisation in terms of the y-deformed Todd polynomials which are formed from the corresponding Chern classes and from the polynomials on y :
\begin{center}
{ \mathversion{bold} ${\chi}_y(T_{X})  = \sum_{j=0}^{\infty}T_j(c_1(T_{X}),...,c_j(T_{X}),y) $ }
\end{center}

The y-Todd  polynomials are given by:
\begin{center}
{ \mathversion{bold} $ T_0(c_0(T_{X}),y) =T _0(1,y)=1 $ }
\end{center}
\begin{center}
{ \mathversion{bold} $ T_1(c_1(T_{X}),y) = \frac{\rm (1-y)c_1(T_{X})}{\rm 2} $ }
\end{center}
\begin{center}
{ \mathversion{bold} $ T_2(c_1(T_{X}),c_2(T_{X}),y) = \frac{\rm (y+1)^2c_1(T_{X})^2+(y^2-10y+1)c_2(T_{X})}{\rm 12} $ }
\end{center}
\begin{center}
{ \mathversion{bold} $ T_3(c_1(T_{X}),c_2(T_{X}),c_3(T_{X}),y) = \frac{\rm -(y+1)^2(y-1)c_1(T_{X})c_2(T_{X})+12y(y-1)c_3(T_{X})}{\rm 24} $ }
\end{center}.

Then one has:

\begin{center}
{ \mathversion{bold} $ {\chi}_y(T_{X}) = 1+\frac{\rm (1-y)c_1(T_{X})}{\rm 2}+\frac{\rm (y+1)^2c_1(T_{X})^2+(y^2-10y+1)c_2(T_{X})}{\rm 12}+\frac{\rm -(y+1)^2(y-1)c_1(T_{X})c_2(T_{X})+12y(y-1)c_3(T_{X})}{\rm 24} $ }
\end{center}.

When X is a Calabi-Yau threefold then the chi-y-genus is given by

\begin{center}
{ \mathversion{bold} $ {\chi}_y(T_{X}) = 1+\frac{\rm (y^2-10y+1)c_2(T_{X})}{\rm 12}+\frac{\rm 12y(y-1)c_3(T_{X})}{\rm 24} $ }
\end{center}.

From this one can to write the following formula for the y-deformed geometric version of the Witten effect:

\begin{center}
{ \mathversion{bold} $\sqrt{{\chi}_y(T_{X})} =[X]+\frac{\rm (y^2-10y+1)c_{2}[X]}{\rm 24}+\frac{\rm y(y-1)c_{3}[X]}{\rm 4} $ }
\end{center}
when y=0 one obtains the usual Witten effect:

\begin{center}
{ \mathversion{bold} $\sqrt{{\chi}_0(T_{X})} =[X]+\frac{\rm (0^2-0+1)c_{2}[X]}{\rm 24}+\frac{\rm 0(0-1)c_{3}[X]}{\rm 4}=[X]+\frac{\rm c_{2}[X]}{\rm 24} $ }
\end{center}

For the other hand  the y-deformed Chern Character $ch_y(G)$  is given by:
$ ch_y(G)= \sum_{i=0}^{3}ch_{i,y}(G)$ with $ ch_{i,y}(G){\in}(H_{6-2i}(X;Q){\otimes}Q[y])$, which can be computed using y-deformed homology-cohomology duality: always one can to have a y-deformed resolution of G by y-deformed locally free sheaves  $(V_{*}) $,in such way that one can to set that $ ch_y(G):= \sum_{i=0}^{3}(-1)^ich_y(V_i)$,and these result does not depend on the choise of the y-deformed resolution.
The total Chern Class for G  has the following sumarization:
\begin{center}
{ \mathversion{bold} $ c(G) = \sum_{j=0}^{q}c_j(G) $ }
\end{center} 
also, the total Chern Class for G has the following factorization:

\begin{center}
{ \mathversion{bold} $ c(G) = \prod_{i=1}^{q}(1+z_i)$ }
\end{center}
The total Chern character of G is defined by:
\begin{center}
{ \mathversion{bold} $ ch(G) = \sum_{j=1}^{q}e^{z_i} $ }
\end{center} 

The total y-deformed Chern character for G  has the following sumarization:

\begin{center}
{ \mathversion{bold} $ ch_y(G) = \sum_{j=1}^{q}e^{(1+y)z_i} $ }
\end{center}
The total y-deformed Chern character for G has the following expantion in terms of the Chern class of G and polynomials for y:

\begin{center}
{ \mathversion{bold} $ ch_y(G) =rk(G)+(y+1)c_{1}(G)+(y+1)^2({\frac{\rm c_{1}(G)^{2}-c_2(G)}{\rm 2}})+(y+1)^3({\frac{\rm c_{1}(G)^{3}-3c_1(G)c_2(G)+3c_3(G)}{\rm 6}})
 $ }
\end{center}

It is easy to see that when y=0, one obtains the usual expantion for the usual Chern character.
For the investigation of the topological aspects of the y-deformed  D-branes is of the great importance to obtain several basic y-deformed invariants of y-deformed BPS D-Branes. One of these y-deformed invariants is the y-deformed RR charge of the y-deformed D-brane. Other y-deformed invariant is the y-deformed  intersection form on y-deformed D-branes on X . This y-deformed invariant for intersections of 
two y-deformed Dp-branes is obtained by multiplication of the y-deformed Mukay vectors of the y-deformed RR charges corresponding to the intersecting y-deformed Dp-branes and is given by: 
 
\begin{center}
\setlength{\baselineskip}{30pt}
{ \mathversion{bold} $ I_{X,y}(G_1,G_2)=[v_{X,y}(G_1)^v.v_{X,y}(G_2)]_X=[(ch(G_1)\sqrt{{\chi}_y(T_{X})})^v.ch(G_2)\sqrt{{\chi}_y(T_{X})}]_X=[ch(G_1)^v.ch(G_2){\chi}_y(T_{X})]_X $ }
\end{center}

where $ [...]_{X,y} $  evaluates the degree of $ (H_{0}(X;Q){\otimes}Q[y]){\cong}(Q{\otimes}Q[y])$ component, and $v^{\vee}$ flips the sign of $(H_{0}(X){\otimes}Q[y]){\oplus}(H_{4}(X){\otimes}Q[y]))$ y-deformed part of the y-deformed  Mukay vector v . In particular, if G itself is locally free, then $ch_{y}(G)^{\vee}=ch_y(G^{\vee})$, where  $ G^{\vee}=Hom_{X}(G,0_X)$ is the y-deformed dual sheaf. Finally is easy to check that: $I_{X,y}(G_1,G_2) =-I_{X,y}(G_2,G_1)$.

On other hand the y-deformed invariant of intersection between y-deformed D-branes is an application of the y-deformed Hirzebruch-Riemann-Roch and for then you can write:

\begin{center}

{ \mathversion{bold} $ I_{X,y}(G_1,G_2)=\sum_{i=0}^{3}(-1)^idimExt^i_{X,y}(G_1,G_2) $ }
\end{center}
For this reason the skew-symmetric property $I_{X,y}(G_1,G_2) =-I_{X,y}(G_2,G_1)$ of the intersection form $I_{X,y}$  for the intersection of two y-deformed Dp-branes may be attributed to the y-deformed Serre duality: $Ext^i_{X,y}(G_1,G_2){\cong}Ext^{3-i}_{X,y}(G_1,G_2)^{\vee}$. Another interesting comentary is that from the y-deformed integrality theorems for diferential and complex manifolds the y-deformed formula H.R.R. is an polynomial on y  and this assures that $I_{X,y}$ takes values in Q[y].

Now let $J_{X,y}{\in}(H_{4}(X;R){\otimes}R[y]) $  be a y-deformed Kahler form on X, whis is here identified with an y-deformed R-extended ample divisor. The y-deformed classical expression of the y-deformed central charge of the y-deformed D-brane G is then given by [1]:
\begin{center}
{ \mathversion{bold} $ Z_{J_{X,y}}^d(G)=-[e^{-J_{X,y}}.v_{X,y}(G)]_{X}=-\sum_{k=0}^{3}{\frac{\rm (-1)^{k}}{\rm k!}[J_{X,y}^k.v_{X,y,k}(G)]_{X}} $ }
\end{center}

where $v_{X,y,k} $ is the $H_{2k}(X){\otimes}Q[y] $ component of $v_{X,y}{\in}(H_{2*}(X;Q){\otimes}Q[y]) $.

In such way we obtain the three y-deformed invariants: y-deformed RR charge,
y-deformed central charge and y-deformed intersections pairings of two y-deformed BPS Dp-branas. With this aid of some algebraic geometry-topology techniques we can to begin the study of topological aspects of y-deformed BPS Dp-branes bounded on a proyective algebraic surface in a Calabi-Yau threefold X. 

\section{y-deformed BPS Dp-branes localized on a surface in a Calabi-Yau
threefold}

Let f be an embedding of a proyective algebraic surface S in a Calabi-Yau threefold X. In the limit of infinite elliptic fiber, the y-deformed BPS Dp-branes  for which the y-deformed central charge remains  finite are those y-deformed BPS Dp-branes which are confined to the algebraic surface S. The physical and topological propertis of the y-deformed BPS D-p-branes  localized on the algebraic surface S then dependen not on the details of the
global model X, but only on the intrinsic y-deformed  geometry of S ans its y-deformed normal bundle $N_{S,y}=N_{S|X,y}$ which is isomorphic to the y-deformed canonical line bundle $K_{S,y}$. In particular, this means that we can compute the y-deformed central charges of y-deformed BPS D-p-branes using y-deformed local mirror symmetry principle on S.

In a elementary physical configuration you have a y-deformed BPS Dp-brane sticking to S.  Such y-deformed  D-brane sticking to S can be described mathematically by a y-deformed $O_{S}-module E$. For this configuration an important y-deformed topological invariant is the y-deformed Euler number of E (the Euler y-polynomial for E) which is defined by  ${\chi}_{y}(E)=\sum_{j=0}^{2}(-1)^ih^i(S,E,y)$, where $h^i(S,E,y)=dim(H^i(S,E))_y$.
For to obtain the y-deformed Euler number of E or the Euler  polynomial of E
the first thing that one needs is the y-deformed Todd class of S or ${\chi}_y  $ class of S:

\begin{center}
{ \mathversion{bold} $ {\chi}_y(T_S) = [S]+\frac{\rm (1-y)c_1(S)}{\rm 2}+\frac{\rm (y+1)^2c_1(S)^2+(y^2-10y+1)c_2(S)}{\rm 12} $ }
\end{center}

this expansion can be writen as:

\begin{center}
{ \mathversion{bold} $ {\chi}_y(T_S) = [S]+\frac{\rm (1-y)c_1(S)}{\rm 2}+{\chi}_{y}(O_S)[pt] $ }
\end{center}
where:

\begin{center}
{ \mathversion{bold} $ {\chi}_{y}(O_S) = [\frac{\rm (y+1)^2c_1(S)^2+(y^2-10y+1)c_2(S)}{\rm 12} ]_S$ }

\end{center}
 The second thing for to do is to apply the y-deformed H.R.R formula, and then one get:

\begin{center}
\setlength{\baselineskip}{40pt}
{ \mathversion{bold} $ {\chi}_{y}(E) = [ch_{y}(E){\chi}_y(T_S)]_S=[ch_{y}(E)([S]+\frac{\rm (1-y)c_1(S)}{\rm 2}+{\chi}_{y}(O_S)]_S=[(rk(E)+(y+1)c_{1}(E)+(y+1)^2({\frac{\rm c_{1}(E)^{2}-c_2(E)}{\rm 2}}))([S]+\frac{\rm (1-y)c_1(S)}{\rm 2}+{\chi}_{y}(O_S)]_S=rk(E){\chi}_{y}(O_S)+[(y+1)^2({\frac{\rm c_{1}(E)^{2}-c_2(E)}{\rm 2}}))+\frac{\rm (y+1)(1-y)c_1(S).c_{1}(E)}{\rm 2}]_S$ }

\end{center}

From the other side, there is y-deformed canonical push-forward homomorphism
$f_*$ from $H_{2*}(S;Q){\otimes}Q[y]$ to $H_{2*}(X;Q){\otimes}Q[y]$, which
maps a y-deformed cycle on S that on X. Also, on can define the y-deformed coherent sheaf $f_{!}E$ on X by extending E by zero to X/S. Now using the y-deformation of the celebrated Grothendieck-Riemman-Roch formula for the embeding f od S in X, one can to relate the y-deformed chern characters of E and $f_{!}E$ as follows:

\begin{center}
{ \mathversion{bold} $ ch_y(f_{!}E) = f_{*}(ch_{y}(E)\frac{\rm 1}{\rm chi_{y}(N_S)}) $ }
\end{center}

Multiplying the boht sides of the y-deformed GRR formula by $\sqrt{{\chi}_{y}(T_X)} $ , one has:

\begin{center}
{ \mathversion{bold} $ ch_y(f_{!}E)\sqrt{{\chi}_{y}(T_X)} = f_{*}(ch_{y}(E)\sqrt{\frac{\rm chi_{y}(T_S))}{\rm chi_{y}(N_S)}
}) $ }
\end{center}

where we have used the y-deformed proyection formula:

 \begin{center}
{ \mathversion{bold} $ f_{*}(a.f^*b) = f_{*}a.b $ }

\end{center}

with  $a{\in}(H_{2*}(S;Q){\otimes}Q[y]),  b{\in}(H_{2*}(X;Q){\otimes}Q[y]) $

and $f^*{chi_{y}}(T_X)={chi_{y}}(T_S).{chi_{y}}(N_S)$  , which follows from the y-deformed short exact sequence of bundles on S: $0 ------> T_S ---->f^*T_X ---->  N_S ----> 0$, combined with the multiplicative property of the chi-y-genus.

Now the y-deformed BPS Dp-brane on a Calabi-Yau threefold X  is represented by G and y-deformed BPS Dp-brane sticking to S can be described by E then one has $G=f_{!}E$ and following formula for the y-deformed Mukai vector of the y-deformed RR charges of $G=f_{!}E$

\begin{center}
{ \mathversion{bold} $ v_{X,y}(f_{!}E)=ch_{y}(f_{!}E)\sqrt{{\chi}_{y}(T_{X})}{\in}(H_{2*}(X;Q){\otimes}Q[y]):={\oplus}_{i=0}^3(H_{2i}(X;Q){\otimes}Q[y])  $ }
\end{center}

The you have:

\begin{center}
{ \mathversion{bold} $ v_{X,y}(f_{!}E)= f_{*}(ch_{y}(E)\sqrt{\frac{\rm chi_{y}(T_S))}{\rm chi_{y}(N_S)}
})=  f_{*}(v_{S,y}(E))$ }
\end{center}

In such way the y-deformed RR charge of the y-deformed BPS Dp-brane represented by E on S regarded as a y-deformed BPS Dp-brane on  X can written in the following intrinsic description (of the y-deformed RR charge on S):

\begin{center}
{ \mathversion{bold} $ v_{S,y}(E)= ch_{y}(E)\sqrt{\frac{\rm chi_{y}(T_S))}{\rm chi_{y}(N_S)}
}=ch_{y}(E)\sqrt{\frac{\rm chi_{y}(T_S))}{\rm chi_{y}(K_S)}
}$ }
\end{center}

The y-deformed gravitational correction factor for S admits the following expansion:

\begin{center}
\setlength{\baselineskip}{30pt}
{ \mathversion{bold} $\sqrt{\frac{\rm chi_{y}(T_S))}{\rm chi_{y}(K_S)}
} = [S]+\frac{\rm (1-y)c_1(S)}{\rm 2}+\frac{\rm (-10y+1+y^2)c_2(S)+3(y-1)^2c_1(S)^2}{\rm 24}  {\in} (H_{2*}(S;Q){\otimes}Q[y])$ }
\end{center}

As a simple exercise one can to compute the y-deformed RR charge of a y-deformed sheaf on S. For this let i: C---> S be an embedding of a smooth genus g algebraic curve in S with the normal bundle $N_{C}=N_{C\S}$. Then from a lin bundle $L_{C}$ on C, one obtains a y-deformed torsion sheaf $i_{!}L_{C}$ on S and $ch_{y}(i_{!}L_{C})$ can be computed from the y-deformed G.R.R. formula:

\begin{center}
\setlength{\baselineskip}{40pt}   
{ \mathversion{bold} $ ch_y(i_{!}L_{C}) = i_{*}(ch_{y}(L_{C})\frac{\rm 1}{\rm chi_{y}(N_C)})=i_{*}((rk(L_{C})+(y+1)c_{1}(L_{C})(1+\frac{\rm (y-1)c_{1}(N_{C})}{\rm 2}))=i_{*}[C]+((y+1)c_{1}(L_{C})+\frac{\rm (y-1)c_{1}(N_{C})}{\rm 2})[pt]=i_{*}[C]+((y+1)deg(L_{C})+\frac{\rm (y-1)deg(N_{C})}{\rm 2})[pt]$ }
\end{center}

where $deg(L):=[c_{1}(L)]_C$ for a line bundle on C.  Then y-deformed RR charge of the y-deformed BPS Dp-brane bounded on S represented by the y-deformed $O_{S}-module$   $i_{!}L_{C}$ can be computed  as follows:

\begin{center}
\setlength{\baselineskip}{40pt}
{ \mathversion{bold} $ v_{S,y}(i_{!}L_{C})= ch_{y}(i_{!}L_{C})\sqrt{\frac{\rm chi_{y}(T_C))}{\rm chi_{y}(K_C)}
}=(i_{*}[C]+((y+1)deg(L_{C})+\frac{\rm (y-1)deg(N_{C})}{\rm 2})[pt])([C]+\frac{\rm (1-y)c_1(C)}{\rm 2})=(i_{*}[C]+((y+1)deg(L_{C})+(1-y)c_1(C))[pt]){\in}{\oplus}(H_{0}(S){\otimes}Q[y])) $ }
\end{center}

I now turn again to intersection pairings of the y-deformed BPS Dp-branes one has the question about the what is the most appropiate intersection   for on y-deformed D-branes on S. Here we will describe only y-deformed candite. 

The  y-deformed candidate uses the intrinsic y-deformed Mukay vector  $v_{S,y} $ and defines a y-deformed symmetric form:

\begin{center}
{ \mathversion{bold} $ I_{S,y}(E_{1},E_{1})= -[v_{S,y}(E_{1})^{v}.v_{S,y}(E_{2})]_S=\frac{\rm r_{1}r_{2}(y^2-10y+1)chi(S)}{\rm 12})+[r_{1}ch_{2}(E_{2})+r_{2}ch_{2}(E_{1})-c_{1}(E_{1}).c_{1}(E_{2})]_{S}$ }
\end{center}

where  $ch(E)=r[S]+c_{1}(E)+ch_{2}(E) $,  ${\chi}(S)=[c_{2}(S)]_S $ Is THE euler number, and  $v_{y}^{\vee}=-v_{0,y}+v_{1,y}-v_{2,y} $ with $v_{i,y}$
being the y-deformed $(H_{2i}(S){\otimes}Q[y]) $ componente of the y-deformed vector $v_{y}$.

In constrast with $I_{X}$ that have values in  $Q[y] $ and when y=0 then takes values in Z, now $I_{S} $ also have values in  $Q[y] $ but in this case when y=0 $I_{S} $ is not Z-valued in general.

\section{References}

\setlength{\baselineskip}{20pt}
[1]   hep-th/0007243

[2]   F. Hirzebruch, Topological Methods in Algebraic Geometry, 1978

\setlength{\baselineskip}{50pt}   
\end{document}